
\magnification \magstep1
\raggedbottom
\openup 2\jot
\voffset6truemm
\headline={\ifnum\pageno=1\hfill \else
\hfill{\it Coulomb Gauge in One-Loop
Quantum Cosmology} \hfill
\fi}
\centerline {\bf COULOMB GAUGE IN ONE-LOOP}
\centerline {\bf QUANTUM COSMOLOGY}
\vskip 1cm
\centerline {\bf Giampiero Esposito$^{1,2,*}$ and Alexander Yu
Kamenshchik$^{3,**}$}
\vskip 1cm
\centerline {\it ${ }^{1}$Istituto Nazionale di Fisica Nucleare}
\centerline {\it Mostra d'Oltremare Padiglione 20, 80125 Napoli, Italy;}
\centerline {\it ${ }^{2}$Dipartimento di Scienze Fisiche}
\centerline {\it Mostra d'Oltremare Padiglione 19, 80125 Napoli, Italy;}
\centerline {\it ${ }^{3}$Nuclear Safety Institute}
\centerline {\it Russian Academy of Sciences}
\centerline {\it 52 Bolshaya Tulskaya, Moscow 113191, Russia.}
\vskip 1cm
\noindent
{\bf Abstract.} The well-known discrepancies between covariant
and non-covariant formalisms in quantum field theory and quantum
cosmology are analyzed by focusing on the Coulomb gauge for vacuum
Maxwell theory. On studying a flat Euclidean background with
boundaries, the corresponding mode-by-mode analysis of one-loop
quantum amplitudes
agrees with the results of the Schwinger-DeWitt technique
and of mode-by-mode calculations in relativistic gauges.
\vskip 4cm
\leftline {$*$E-mail address: esposito@na.infn.it
$\; \; \; \;$ $**$E-mail address: grg@ibrae.msk.su}
\vskip 10cm
The quantization program for gauge fields and gravitation in the
presence of boundaries is receiving careful consideration in the recent
literature [1-12]. The motivations for this analysis come in part from
the quantization of closed cosmologies [13-14],
and in part from the need to
understand the relation between different quantization techniques in
field theory [15-16]. The choices to be
made are as follows: (i) quantization
technique; (ii) background 4-geometry; (iii) boundary 3-geometry;
(iv) boundary conditions respecting BRST invariance and local supersymmetry;
(v) gauge condition; (vi) regularization technique.
For a given choice of field theory, background and corresponding
boundary 3-geometry, different quantization techniques and different
gauge conditions have led to discrepancies for the semi-classical
evaluation of quantum amplitudes [1-12].
This calculation has been performed
within the framework of $\zeta$-function regularization, where the
$\zeta(0)$ value yields both the scaling properties of the amplitudes
and one-loop divergences of physical theories. On reducing a field theory
with first-class constraints to its physical degrees of freedom
{\it before} quantization [1-12],
the resulting $\zeta(0)$ values disagree with
the Schwinger-DeWitt $A_{2}$ coefficient [17-18]. This occurs both for
compact Riemannian 4-manifolds without
boundary [19-21] and for Riemannian
backgrounds with boundary [1-12]. Moreover, further discrepancies have been
found on studying the quantum theory of spin-$1/2$ fields at one-loop
about background 4-geometries with boundaries [5,22-25].

It therefore seems that the discrepancies found in the literature have
at least two origins, as follows.
\vskip 0.3cm
\noindent
(i) A geometrical source, owed to the
singularity at the origin for manifolds with just one boundary. By this
we mean that there is no regular vector field inside matching
the normal at the boundary. Hence the normal and tangential
components of physical fields inside are ill-defined, and it is
impossible to achieve a consistent 3+1 split. Analogously, such a
split cannot be obtained on compact 4-manifolds without boundary,
if their Euler number does not vanish.
\vskip 0.3cm
\noindent
(ii) A field-theoretical source, i.e. non-physical degrees
of freedom and ghost modes yield contributions to $\zeta(0)$ which do
not cancel each other on considering curved backgrounds and/or the presence
of boundaries.

For example, for manifolds with just one boundary, the
mode-by-mode analysis of Faddeev-Popov amplitudes,
which relies on the expansion in harmonics of the potential
(e.g. eqs. (4)-(5)) and hence on the eigenvalue equations for
the modes multiplying each harmonic, disagrees with the
Schwinger-DeWitt evaluation of the same amplitudes and is
gauge-dependent. Moreover, on
manifolds with two boundaries admitting a consistent 3+1 split, the
physical degrees of freedom of vacuum Maxwell theory,
i.e. the transverse part of the potential, do not enable one
to recover the full $\zeta(0)$ value [26]. These calculations, when
combined with a previous analysis of discrepancies for
spin-${1 \over 2}$ fields [24-25], have motivated a mode-by-mode analysis
of Euclidean Maxwell theory and linearized
gravity [26-27] within the framework
of Faddeev-Popov formalism [28-29] on manifolds with two boundaries.
Remarkably, in this case the Schwinger-DeWitt and mode-by-mode
formalisms for BRST-covariant Faddeev-Popov amplitudes
are found to agree. While our previous analysis focused
on relativistic gauges for such theories, it appears necessary to
complete this investigation by considering non-relativistic gauges as well.
Hence we here study the Coulomb gauge for vacuum Maxwell theory about
flat Euclidean backgrounds with two boundaries, since such a gauge choice
is more relevant for reduction to physical degrees of freedom and Hamiltonian
formalism [30].

For this purpose, we use the version of $\zeta$-function technique
[9,31] elaborated in refs. [6-8]. Following refs. [6-8], we write
$f_{n}(M^{2})$ for the function occurring in the equation obeyed by
the eigenvalues by virtue of boundary conditions, and $d(n)$ for the
degeneracy of the eigenvalues parametrized by the integer $n$. One then
defines the function
$$
I(M^{2},s) \equiv \sum_{n=n_{0}}^{\infty} d(n) \; n^{-2s}
\; \log f_{n}(M^{2}) \; .
\eqno (1)
$$
Such a function has an analytic continuation to the whole complex-$s$
plane as a meromorphic function, i.e.
$$
I(M^{2},s)={I_{\rm pole}(M^{2})\over s}+I^{R}(M^{2})
+O(s) \; .
\eqno (2)
$$
The $\zeta(0)$ value is then obtained as [6-8]
$$
\zeta(0)=I_{\rm log}+I_{\rm pole}(\infty)-I_{\rm pole}(0)
\; ,
\eqno (3)
$$
where $I_{\rm log}$ is the coefficient of $\log \; M$ from
$I(M^{2},s)$ as $M \rightarrow \infty$, and $I_{\rm pole}(M^{2})$
is the residue at $s=0$.

To perform the one-loop analysis of vacuum Maxwell theory, we
expand the normal and tangential components of the electromagnetic
potential on a family of 3-spheres. With the notation of [2,9-10,26],
one writes
$$
A_{0}(x,\tau)=\sum_{n=1}^{\infty} R_{n}(\tau) \; Q^{(n)}(x) \; ,
\eqno (4)
$$
$$
A_{k}(x,\tau)=\sum_{n=2}^{\infty} \biggr[f_{n}(\tau) \; S_{k}^{(n)}(x)
+g_{n}(\tau) \; P_{k}^{(n)}(x) \biggr] \; .
\eqno (5)
$$
Of course, $Q^{(n)}(x), S_{k}^{(n)}(x)$ and $P_{k}^{(n)}(x)$ are
scalar, transverse and longitudinal harmonics
on $S^{3}$ respectively [1-2,32].

Within the Faddeev-Popov quantization
scheme [10,28-29], after adding to the
Euclidean Lagrangian the Coulomb gauge-averaging term
${1\over 2\alpha} \tau^{-4} {\biggr(A_{i}^{\; \; \mid i}\biggr)}^{2}$,
one finds eigenvalue equations on taking variations of the total
Euclidean action with respect to the modes $f_{n},g_{n},R_{n}$ of
eqs. (4)-(5) [10,26]. If $\alpha=1$, these equations take the following
form (the decoupled mode will be treated separately):
$$
\left({d^{2}\over d\tau^{2}}+{1\over \tau}{d\over d\tau}
-{n^{2}\over \tau^{2}}+\lambda_{n}\right)f_{n}(\tau)=0 \; ,
\eqno (6)
$$
$$
{\widehat A}_{n} g_{n}(\tau)+{\widehat B}_{n} R_{n}(\tau)=0 \; ,
\eqno (7)
$$
$$
{\widehat C}_{n} g_{n}(\tau)+{\widehat D}_{n} R_{n}(\tau)=0 \; ,
\eqno (8)
$$
where [33]
$$
{\widehat A}_{n} \equiv {d^{2}\over d\tau^{2}}
+{1\over \tau}{d\over d\tau}-{(n^{2}-1)\over \tau^{2}}
+\lambda_{n} \; ,
\eqno (9)
$$
$$
{\widehat B}_{n} \equiv -(n^{2}-1) \biggr({d\over d\tau}
+{1\over \tau} \biggr) \; ,
\eqno (10)
$$
$$
{\widehat C}_{n} \equiv {1\over \tau^{2}} {d\over d\tau} \; ,
\eqno (11)
$$
$$
{\widehat D}_{n} \equiv -{(n^{2}-1)\over \tau^{2}}
+\lambda_{n} \; .
\eqno (12)
$$
Moreover, in the Coulomb gauge, the ghost eigenvalue equation is
found to be
$$
{(n^{2}-1)\over \tau^{2}} \epsilon_{n}(\tau)=\lambda_{n} \;
\epsilon_{n}(\tau) \; \; \; \;
\forall n \geq 1 \; ,
\eqno (13)
$$
where the modes $\epsilon_{n}$ are the ones occurring in the
expansion on a family of 3-spheres of the scalar field
$\epsilon(x,\tau)$ as $\sum_{n=1}^{\infty}\epsilon_{n}(\tau) \;
Q^{(n)}(x)$. Their contribution to $\zeta(0)$ is finally
multiplied by -2 [9-10,26].

We here use magnetic boundary conditions
[3,10,26]. Hence $A_{k}(x,\tau)$
is set to zero at the 3-sphere boundaries as well as the Coulomb
gauge $\Phi_{c}(A) \equiv \tau^{-2}A_{i}^{\; \; \mid i}$, and
correspondingly the ghost modes
$\epsilon_{n}$ [10,26].

The solution of eq. (6) is $f_{n}=A_{n}I_{n}(M\tau)
+B_{n} K_{n}(M\tau)$, where $M^{2}=-\lambda_{n}$. By virtue of
our boundary conditions, such a solution should vanish at
$\tau=\tau_{-}$ and at $\tau=\tau_{+}$, where $\tau_{-},\tau_{+}$
are the radii of the two concentric 3-sphere boundaries. Hence the
corresponding eigenvalue condition can be written as
$$
I_{n}(M\tau_{-})K_{n}(M\tau_{+})
-I_{n}(M\tau_{+})K_{n}(M\tau_{-})=0 \; .
$$
By using the technique described in refs. [6-8] and outlined in
eqs. (1)-(3), one can show that the contribution of transverse
modes to $\zeta(0)$ is [26]
$$
\zeta_{\rm phys}(0)=-{1\over 2} \; .
\eqno (14)
$$

{}From eq. (13) one can see that the only non-vanishing contribution
from ghost modes is obtained when $n=1$, which implies that
$\lambda_{n}=0$.
The corresponding eigenfunctions are {\it arbitrary}
functions of $\tau$ which obey homogeneous Dirichlet conditions at
$\tau=\tau_{-}$ and at $\tau=\tau_{+}$. The general eigenfunction
can be written as
$$
\epsilon_{1}(\tau) = \sum_{k=1}^{\infty} c_{k} \sin\left(
{\pi k (\tau - \tau_{-}) \over (\tau_{+} - \tau_{-})}\right) \; .
\eqno (15)
$$
Note that we here face an entirely new situation, in that we have
to deal with an infinite number of zero-modes.
The inclusion of zero-modes into the general expression
for $\zeta(0)$ was studied in
mathematical papers [34] and for the
problems of quantum gravity as well
[35-36]. It is known that when we
have a {\it finite} number of zero-modes we have simply to add it to
the $\zeta(0)$ value, however, we do not know {\it a priori} what should
one do with an infinite number of such modes.
Hence we need the appropriate regularization
of an infinite number of our zero-modes.
We try to achieve this being inspired by the ideas of $\zeta$-
regularization technique. For this purpose, we point out
that the eigenfunctions (15) belong to the space
whose elements can be parametrized by the natural numbers $k=1,2,...$.
All these eigenfunctions can be treated on equal footing. Thus, we
are led to define the {\it regularized}
dimension of this space as the regularized number
of its basis elements, or, within the framework of $\zeta$-function
technique (see below),
as $\zeta_{R}(0) = -{1 \over 2}$, where $\zeta_{R}(s)
\equiv \sum_{n=1}^{\infty} n^{-s}$ is the usual Riemann $\zeta$-
function, whose properties and values are well-known [37]. Thus
$$
\zeta_{\rm ghost}(0) = -{1 \over 2} \; .
\eqno (16)
$$
Of course, {\it infinitely many} definitions of regularized dimension
are possible on considering the zeta-functions
$$
\zeta_{R,a}(s) \equiv \sum_{n=1}^{\infty}(n+a)^{-s} \; ,
$$
where $a$ is a real parameter. Our choice corresponds to
the value $a=0$, and it appears plausible to
take into account that the sine functions of eq. (15) are
labelled by the integers $k=1,2,3...$ only.

By virtue of our boundary conditions, coupled gauge modes vanish
at the boundaries, i.e. $g_{n}(\tau_{-})=g_{n}(\tau_{+})=0,
R_{n}(\tau_{-})=R_{n}(\tau_{+})=0$. The boundary conditions
$$
R_{n}(\tau_{+})=R_{n}(\tau_{-})=0 \; \; \; \;
\forall n \geq 1
$$
can be obtained from a non-trivial application of gauge
invariance. In other words, one may start from a relativistic
gauge condition written in the form (cf. [33])
$$
\Phi(A) \equiv \lambda \; A_{0} \; {\rm Tr} \; K
+ { }^{(3)}\nabla^{i}A_{i} \; ,
\eqno (17)
$$
where $\lambda$ is a dimensionless parameter and
${ }^{(3)}\nabla^{i}A_{i}=\tau^{-2}A_{i}^{\; \; \mid i}$. Now it
is well-known that magnetic boundary conditions in relativistic
gauges $\Phi_{R}$ imply that any such $\Phi_{R}$ should vanish
at the boundaries [26]. When combined with homogeneous Dirichlet
conditions on the tangential components of the electromagnetic
potential, which require that
$$
f_{n}(\tau_{+})=f_{n}(\tau_{-})=0 \; , \;
g_{n}(\tau_{+})=g_{n}(\tau_{-})=0 \; , \;
\forall n \geq 2 \; ,
$$
this implies that $A_{0}(x,\tau_{+})=A_{0}(x,\tau_{-})=0$,
i.e.
$$
R_{n}(\tau_{+})=R_{n}(\tau_{-})=0
\; \; \; \;
\forall n \geq 1 \; .
$$
If gauge invariance is respected in quantum theory, this set of conditions
holds for all values of $\lambda$.
Hence, on taking the limit as
$\lambda \rightarrow 0$ in (17), one recovers the Coulomb gauge we are
interested in, subject to the {\it same} boundary conditions on
$g$-modes and $R$-modes, {\it providing} gauge invariance holds.
The legitimacy of this procedure depends crucially on a
{\it direct} proof of gauge invariance in the presence of boundaries,
not relying on the formal arguments frequently presented in the
literature, and is the object of a paper in preparation by ourselves
and other co-authors [33]. We are then able to obtain, in particular,
boundary conditions on the decoupled mode $R_{1}$, which would otherwise
remain {\it totally} arbitrary.

Moreover, eqs. (7)-(12) imply
that
$$
R_{n}(\tau)={{\dot g}_{n}(\tau) \over [(n^{2}-1)+M^{2}\tau^{2}]} \;,
\eqno (18)
$$
where $g_{n}(\tau)$ obeys the second-order differential equation
$$ \eqalignno{
\; & \Bigr[(n^{2}-1)+M^{2}\tau^{2}\Bigr]M^{2}\tau^{2}
\; {d^{2}g_{n}\over d\tau^{2}}
+\Bigr[3(n^{2}-1)+M^{2}\tau^{2}\Bigr]M^{2}\tau \;
{dg_{n}\over d\tau} \cr
&-{\Bigr[(n^{2}-1)+M^{2}\tau^{2}\Bigr]}^{2} \;
{(n^{2}-1)\over \tau^{2}} g_{n} \cr
&-M^{2} {\Bigr[(n^{2}-1)+M^{2}\tau^{2}\Bigr]}^{2}g_{n}=0 \; .
&(19)\cr}
$$
Note that, if $M=0$, the limiting form of (19) is
$$
{(n^{2}-1) \over \tau^{2}} g_{n} = 0
\; \; \; \; \forall n\geq 2 \; ,
$$
whose only solution is $g_{n}(\tau)=0, \forall n \geq 2$
and $\forall \tau \in [\tau_{-},\tau_{+}]$.
If $M$ does not vanish, we may regard (19) as a second-order
ordinary differential equation.
Remarkably, one then deals with an overdetermined
problem, in that both $g_{n}$ and ${\dot g}_{n}$ have to vanish
at the boundaries in the light of boundary conditions and of
eq. (18). Hence the only solution is the trivial one, i.e.
$g_{n}(\tau)=0 \; \forall \tau \in [\tau_{-},\tau_{+}]$
and $\forall n \geq 2$, and similarly for $R_{n}(\tau), \forall
n \geq 2$. Thus, coupled gauge modes give a vanishing contribution to
the full $\zeta(0)$:
$$
\zeta_{\rm coupled}(0) = 0 \; .
\eqno (20)
$$

By contrast, the decoupled mode $R_{1}(\tau)$ can be represented by
an arbitrary function vanishing at $\tau=\tau_{-}$ and at $\tau=\tau_{+}$.
Hence, one can apply again the argument leading to eq. (16), which implies
$$
\zeta_{R_{1}}(0)=-{1 \over 2} \; .
\eqno (21)
$$
By virtue of eqs. (14), (16), (20)-(21) one finds
$$
\zeta(0)=\zeta_{\rm phys}(0)+\zeta_{\rm coupled}(0)
+\zeta_{R_{1}}(0)-2\zeta_{\rm ghost}(0) = 0 \; .
\eqno (22)
$$

One can also consider another approach to boundary
conditions for the electromagnetic field
subject to the Coulomb gauge. For this purpose, we can view the
normal component $A_{0}$ of the electromagnetic field as a
Lagrange multiplier which should not be included in the gauge conditions
and which should be integrated over not only inside the manifold
under consideration, but also on its boundaries [30].
In this case the homogeneous mode $R_{1}$ should be excluded,
since it does not correspond to any constraint
(really, the constraint $\nabla_{i}F^{i0}$ of the electromagnetic
field has
no homogeneous modes). Moreover, homogeneous ghost modes should also
be discarded, because no homogeneous modes correspond
to the Coulomb gauge condition (the problem of discarding ghost
zero-modes was discussed in a slightly different framework in [38]).

Thus, in such an approach to the problem of boundary
conditions and the treatment of Lagrange multipliers and ghost modes,
there are no non-zero contributions to the full $\zeta(0)$ from the
decoupled mode and from ghost zero-modes (cf. eqs. (16), (21)). However,
in this case we have a non-trivial contribution from the coupled
gauge modes. Hence our second-order equation (19) is no longer
overdetermined, since there are no boundary conditions
on $R_{n}$ (see (18)), and we have a Dirichlet boundary-value
problem involving $g_{n}$-modes only.
One can then show that the basis function of eq. (19) can be
represented as
$$
g_{n}(\tau) = w_{\nu}(M \tau)\;v_{n}(M \tau)\;x_{n}(M,\tau) \; ,
\eqno (23)
$$
where $w_{\nu}$ is a linear combination of modified Bessel functions
$I_{\nu}$ and $K_{\nu}$ with $\nu^{2} \equiv 2(n^{2}-1)$. Then
$$
v_{n}(M \tau) = \sqrt{{[(n^{2}-1)+M^{2}\tau^{2}] \over M^{2}\tau^{2}}}
$$
and $x_{n}(M,\tau)$ obeys the equation
$$
w_{\nu}\left[{\partial^{2} x_{n} \over \partial \tau^{2}}
+ {1 \over \tau}\; {\partial x_{n} \over \partial \tau}
+ F(n,M,\tau) x_{n} \right]+2{dw_{\nu}\over d\tau}
{\partial x_{n} \over \partial \tau}=0 \; ,
\eqno (24)
$$
where
$$ \eqalignno{
F(n,M,\tau) & \equiv -{(n^{2}-1)\over M^{2} \tau^{4}}
\left[(n^{2}-1)+
{\Bigr(1+{(n^{2}-1) \over M^{2} \tau^{2}} \Bigr)}^{-1}
\biggr({3(n^{2}-1)+M^{2}\tau^{2} \over
{(n^{2}-1)+M^{2}\tau^{2}}} \biggr) \right] \cr
&-{(n^{2}-1)\over M^{2}}
{\Bigr(1+{(n^{2}-1) \over M^{2} \tau^{2}} \Bigr)}^{-1} \times \cr
& \times \left[-{3\over \tau^{4}}
+{(n^{2}-1)\over M^{2} \tau^{6}}
{\Bigr(1+{(n^{2}-1) \over M^{2} \tau^{2}} \Bigr)}^{-1}
\right] \; .
&(25)\cr}
$$
The solution of (24) has the following asymptotic forms:
$$
x_{n}(M,\tau) \sim A_{n} +{B_{n}(\tau)\over M^{2}}
\; {\rm as} \; M \rightarrow \infty \; ,
$$
where $A_{n}$ is a constant, and
$$
x_{n}(M,\tau) \sim \exp \left[-{(n^{2}-1)\over M\tau}\right]
\; {\rm as} \; M \rightarrow 0 \;.
$$
Hence one can show that the only function
in the product (23) which can
give a non-trivial contribution to $\zeta(0)$ is $w_{\nu}$, and its
contribution can be easily calculated along the lines described in
refs. [6-8,24-27] and applied above to the calculation of the
contribution of physical modes (eq. (14)). This leads to
$$
\zeta_{\rm coupled}(0) = {1 \over 2}\;.
\eqno (26)
$$
Combining eqs. (14) and (26) one has again
$$
\zeta(0) = \zeta_{\rm phys}(0) + \zeta_{\rm coupled}(0)
=0 \; .
$$
Thus, our result obtained with another treatment of boundary
conditions coincides with the one resulting from eq. (22).

Interestingly, our result for the full $\zeta(0)$ agrees with previous
calculations for relativistic gauges [26] and with the $A_{2}$ coefficients
of the Schwinger-DeWitt technique [26]. For our manifold with two
3-sphere boundaries, the general formulae for the Schwinger-DeWitt
$A_{2}$ coefficient [3,5,18] yield zero, since the contributions of the
two boundaries cancel each other.

We should emphasize that part of our analysis relies on the
Riemann-$\zeta$ regularization of the infinite dimension
of the space of zero-modes for the decoupled component of
$A_{0}$ and for the ghost field. We can only say that this
procedure leads to the remarkable cancellation occurring in
(22), but we cannot as yet put it on more rigorous grounds.
Moreover, our boundary conditions in the Coulomb gauge result,
in the first approach,
from a non-trivial limiting procedure out of the well-established
set of boundary conditions imposed on studying relativistic
gauges. Such a limiting procedure leads to a non-vanishing
contribution to $\zeta(0)$ resulting from ghost modes. As it is
well-known, this non-vanishing contribution is absent in the
case of flat space without boundaries.
The ultimate justification of our approach can be obtained by proving
in a direct way the gauge invariance of quantum amplitudes
for problems with boundaries, rather than assuming it. A
detailed analysis of such a problem can be found in ref. [33].
It seems encouraging, however, that boundary conditions for the
Coulomb gauge derived from the familiar, non-relativistic
framework, lead to a $\zeta(0)$ value equal to the one in (22).

In agreement with recent work by the authors [26-27], our
analysis of the Coulomb gauge adds evidence in favour of gauge-
and ghost-modes being necessary to obtain gauge-invariant
amplitudes for manifolds with boundaries. In general, there is no
exact cancellation of such contributions on non-trivial backgrounds.
Moreover, we should say that a very recent paper by Moss and Poletti
[39], relying on previous work by Vassilevich [40], has corrected
all previous Schwinger-DeWitt calculations for spin$>0$ for
manifolds with boundaries. Hence they find full agreement between
our investigations of the conformal anomalies for
spin-${1\over 2}$ and spin-$1$ fields [22-26], and their corrected
analysis, even in the 1-boundary case. However, the problem
remains of understanding why, in the 1-boundary case only, the
mode-by-mode evaluation of one-loop quantum amplitudes is
gauge-dependent, as shown in [26]. As far as we can see, this last
remaining discrepancy seems to point out to some serious limitations
which apply when the background 4-geometry does not admit a
well-defined 3+1 decomposition.

The problem of the correct choice of
technique for one-loop calculations in quantum cosmology is also
important when one investigates
the properties of the wave function of the universe [13-14].
These properties are crucial for studying the relation
between quantum gravity, inflationary
cosmology and particle physics [41-42].
\vskip 0.3cm
\leftline {\bf Acknowledgement}
\vskip 0.3cm
The research described in this publication was made possible in part
by grant No MAE000 from the International Science Foundation.
A. Kamenshchik is grateful to the Dipartimento di Scienze Fisiche
dell'Universit\`a di Napoli and to the Istituto Nazionale di
Fisica Nucleare for kind hospitality and financial support during his
visit to Naples in May 1994. The work of A. Kamenshchik was
partially supported  by the Russian Foundation for Fundamental
Researches through grant No 94-02-03850-a.
Moreover, our joint paper was supported in part by the European
Union under the Human Capital and Mobility Program.
Collaboration with
I.V. Mishakov and G. Pollifrone on related topics [33] is gratefully
acknowledged by both authors. Last, but not least, enlightening
correspondence with A.O. Barvinsky has greatly stimulated
and improved our research.
\vskip 0.3cm
\leftline {\bf References}
\vskip 0.3cm
\item {[1]}
K. Schleich, Phys. Rev. D 32 (1985) 1889.
\item {[2]}
J. Louko, Phys. Rev. D 38 (1988) 478.
\item {[3]}
I.G. Moss and S.J. Poletti, Nucl. Phys. B 341 (1990) 155.
\item {[4]}
I.G. Moss and S.J. Poletti, Phys. Lett. B 245 (1990) 355;
S.J. Poletti, Phys. Lett. B 249 (1990) 249.
\item {[5]}
H.C. Luckock, J. Math. Phys. 32 (1991) 1755.
\item {[6]}
A.O. Barvinsky, A.Yu. Kamenshchik, I.P. Karmazin and I.V. Mishakov,
Class. Quantum Grav. 9 (1992) L27.
\item {[7]}
A.O. Barvinsky, A.Yu. Kamenshchik and I.P. Karmazin,
Ann. Phys. (N.Y.) 219 (1992) 201.
\item {[8]}
A.Yu. Kamenshchik and I.V. Mishakov, Int. J. Mod. Phys. A 7
(1992) 3713.
\item {[9]}
G. Esposito, Quantum gravity, quantum cosmology and
Lorentzian geometries, Lecture Notes in Physics, new
series m: Monographs, vol. m12, second corrected and
enlarged edition (Springer-Verlag, Berlin, 1994).
\item {[10]}
G. Esposito, Class. Quantum Grav. 11 (1994) 905.
\item {[11]}
G. Esposito, Nuovo Cimento B 109 (1994) 203.
\item {[12]}
G. Esposito, Int. J. Mod. Phys. D (1994) (to be published).
\item {[13]}
J.B. Hartle and S.W. Hawking, Phys. Rev. D 28 (1983) 2960.
\item {[14]}
S.W. Hawking, Nucl. Phys. B 239 (1984) 257.
\item {[15]}
M. Henneaux, Phys. Rep. 126 (1985) 1.
\item {[16]}
M. Henneaux and C. Teitelboim, Quantization of gauge systems
(Princeton University Press, Princeton, 1992).
\item {[17]}
B.S. DeWitt, Dynamical theory of groups and fields
(Gordon and Breach, New York, 1965).
\item {[18]}
T.P. Branson and P.B. Gilkey, Commun. Part. Diff. Eqns.
15 (1990) 245.
\item {[19]}
P.A. Griffin and D.A. Kosower, Phys. Lett. B 233 (1989) 295.
\item {[20]}
D.V. Vassilevich, Nuovo Cimento A 104 (1991) 743.
\item {[21]}
D.V. Vassilevich, Nuovo Cimento A 105 (1991) 649.
\item {[22]}
P.D. D'Eath and G. Esposito, Phys. Rev. D 43 (1991) 3234.
\item {[23]}
P.D. D'Eath and G. Esposito, Phys. Rev. D 44 (1991) 1713.
\item {[24]}
A.Yu. Kamenshchik and I.V. Mishakov, Phys. Rev. D 47 (1993) 1380.
\item {[25]}
A.Yu. Kamenshchik and I.V. Mishakov, Phys. Rev. D 49 (1994) 816.
\item {[26]}
G. Esposito, A.Yu. Kamenshchik, I.V. Mishakov and
G. Pollifrone, Euclidean Maxwell theory in the presence of
boundaries, part II, DSF preprint 94/4, March 1994.
\item {[27]}
G. Esposito, A.Yu. Kamenshchik, I.V. Mishakov and
G. Pollifrone, Gravitons in one-loop quantum cosmology:
correspondence between covariant and non-covariant formalisms,
DSF preprint 94/14, May 1994.
\item {[28]}
L.D. Faddeev and V. Popov, Phys. Lett. B 25 (1967) 29.
\item {[29]}
L. D. Faddeev, Theor. Math. Phys. 1 (1969) 1.
\item {[30]}
A.O. Barvinsky, Phys. Rep. 230 (1993) 237.
\item {[31]}
S.W. Hawking, Commun. Math. Phys. 55 (1977) 133.
\item {[32]}
E.M. Lifshitz and I.M. Khalatnikov, Adv. Phys. 12 (1963) 185.
\item {[33]}
G. Esposito, A.Yu. Kamenshchik, I.V. Mishakov and
G. Pollifrone, Gauge invariance in quantum cosmology,
work in progress, July 1994.
\item {[34]}
M.F. Atiyah, V.K. Patodi and I.M. Singer, Math. Proc.
Camb. Phil. Soc. 79 (1976) 71.
\item {[35]}
S.M. Christensen and M.J. Duff, Nucl. Phys. B 170 (1980) 480.
\item {[36]}
E.S. Fradkin and A.A. Tseytlin, Nucl. Phys. B 234 (1984) 472.
\item {[37]}
M. Abramowitz and I. Stegun, Handbook of mathematical functions
with formulas, graphs and mathematical tables, Natl. Bur. Stand.
Appl. Math. Ser. No 55 (U.S. GPO, Washington, D.C., 1965).
\item {[38]}
T.R. Taylor and G. Veneziano, Nucl. Phys. B 345 (1990) 210.
\item {[39]}
I.G. Moss and S.J. Poletti, Phys. Lett. B 333 (1994) 326.
\item {[40]}
D.V. Vassilevich, Vector fields on a disk with mixed boundary
conditions, St. Petersburg preprint SPbU-IP-94-6.
\item {[41]}
A.O. Barvinsky and A.Yu. Kamenshchik, Class. Quantum. Grav.
7 (1990) L181.
\item {[42]}
A.Yu. Kamenshchik, Phys. Lett. B 316 (1993) 45.
\bye